\documentclass{appolb}
\usepackage{epsfig}
\begin{document}
\preprint{FR-PHENO-2012-19.}
\title{A simple $SU(5)$ model with unification near the Planck scale.%
%\thanks{Presented at ...}%
}
\author{J.~J.~van~der~Bij
\address{Institut f\"ur Physik,\\
 Albert-Ludwigs-Universit\"at Freiburg,\\
 79104 Freiburg, Germany}
}
\maketitle
\begin{abstract}
We study unification in the $SU(5)$ model with an extra Dirac multiplet
in the ${\bf 24}$ representation. After spontaneous symmetry breaking
we have at low energies a singlet, a colorless triplet and a neutral 
color-octet. All other particles can be taken at the unification scale.
This combination leads to a unification very near the Planck scale.
The triplet is light, its neutral component is a dark matter candidate.
The model is in agreement with a recently derived anomaly condition,
that implies that the number of (Weyl)-fermions has to be a multiple of
16. 
\end{abstract}
  
\section{Introduction}
The spectrum of known particles leads one to consider the unification of the known gauge
forces into a larger group. The simplest one is the gauge group $SU(5)$. 
Larger groups, like $SO(10)$ or $E(6)$ are also popular. However the
known particles do not lead to a unification of the gauge couplings. Therefore one needs to 
add new particles to the theory. Also dark matter exists, which is most easily explained by the
presence of new particles around the electroweak scale. One is therefore interested in 
finding a suitable set of new particles to get a consistent view of unification.
Very popular is the extension towards a supersymmetric model. However this extension is
not completely without problems.
Let us set out what would be desirable for a successful description of unification.
We can consider the following conditions.
\begin{enumerate}
\item The gauge coupling constants should unify at a high scale.
\item Preferably the scale should be very near the Planck scale, in order to have a link
      with gravity.
\item The extra particles should not lead to proton decay.
\item There should be a dark matter candidate.
\item There should not be extra interactions in conflict with phenomenology.
In particular flavor changing neutral currents are dangerous.
\item Preferably there should be an independent reason for the choice of gauge
 group and representations.
\end{enumerate}
The supersymmetric model works very well regarding point one. Doubling the 
standard model particles with their supersymmetric partners leads quite precisely to
a unification within $SU(5)$. On the other points it does not do so well.
The unification scale is below the Planck scale, so a unification with gravity 
is somewhat difficult to imagine. Proton decay has to be prevented by extra symmetries and/or
fine-tuning of parameters. A dark matter candidate can be constructed by the imposition of R-parity,
which however is an extra principle unrelated to supersymmetry itself. One has to 
finely tune a large number of parameters to stay out of conflict with experiment.
So the balance is not purely positive.
Given the fact, that the LHC is starting to constrain the theory quite severely\cite{ichep},
one should keep an open mind and look at possibilities outside supersymmetric
unification.
In particular one would like to  take into account point six in the above list as well.
Normally this is ignored and one takes  the structure of the known particles as motivation only.
However recently some progress in this question has been made\cite{vdbij1,vdbij2}. On the basis
of a gravitational anomaly within a cosmological context,  some constraints
on the possible existence of particle multiplets were found. More precisely the indication is
that the number of gauge fields has to be a multiple of 8 and the number of (Weyl)-fermions
a multiple of 16. This fits in well with the known particle content and with the gauge group
$SU(5)$, however without supersymmetry.\

\section{The model}

Given the above we therefore will consider a $SU(5)$ unification without supersymmetry.
A promising model was described in \cite{perez1}. The main feature of this model for our 
purpose
is the addition of an extra set of fermions in the representation ${\bf 24}$ of $SU(5)$.
Under the subgroup $SU(3) \times SU(2) \times U(1)$ this representation is decomposed as:
\begin{eqnarray}
{\bf 24} = &(\rho_8) \oplus (\rho_3) \oplus (\rho_{(3,2)}) \oplus (\rho_{(\bar 3, 2)}) \oplus (\rho_0) \\
            \nonumber
         = &({\bf 8,1,0}) \oplus ({\bf 1,3,0}) \oplus ({\bf 3,2, -5/6})\oplus ({\bf 3,2, 5/6})
        \oplus({\bf 1,1,0})
\end{eqnarray}
In a first stage of symmetry-breaking these particles receive a mass through a coupling
to a ${\bf 24_{H}}$ Higgs field and also have a direct mass term. The Lagrangian density is:
\begin{equation}
{\cal L}_{mass, Yukawa} = M\, Tr({\bf 24}^2) + \lambda \, Tr({\bf 24}^2 \, {\bf 24_H}) + h.c.
\end{equation}
The masses derived from this Lagrangian can be written as \cite{perez1}:
\begin{equation}
M_0= \mid M -\Lambda \mid,\,
M_3= \mid M -3 \Lambda \mid,\,
M_8= \mid M + 2\Lambda \mid,\,
M_{32}= \mid M - 1/2 \Lambda \mid
\end{equation}
Here $\Lambda$ is a complex number coming from the product of the Yukawa coupling and the vacuum expectation
 value of the ${\bf 24}$ Higgs field.
Using also extra Higgses that were introduced in order to have a seesaw mechanism for neutrino masses
a unification was possible, however rather complicated, relying strongly on having certain Higgses to be
light. Naively one would like to have these Higgs fields to be at  the unification scale.
Ignoring the presence of the Higgs fields, assuming that they have a mass at the unification scale,
the one-loop renormalization group equations read:

\begin{eqnarray}
2\pi(\alpha_1^{-1}-\alpha_U^{-1}) &=&\frac{41}{10} \ln(m_U/m_Z) + \frac{10}{3} \ln(m_U/m_{32}) \\
                                    \nonumber
2\pi(\alpha_2^{-1}-\alpha_U^{-1}) &=&-\frac{19}{6} \ln(m_U/m_Z) + \frac{4}{3} \ln(m_U/m_3) + 
2 \ln(m_U/m_{32})\\ \nonumber
2\pi(\alpha_3^{-1}-\alpha_U^{-1}) &=&-7 \ln(m_U/m_Z) 
+ 2 \ln(m_U/m_8) + \frac{4}{3} \ln(m_U/m_{32})
\end{eqnarray}

In these formulas $m_U$ and $\alpha_U$ are the unification scale and the unified fine structure constant
at the unification scale.
Given the formula (3) for the masses  a sensible unification in this form is not possible,
without adding light scalars into the renormalization group running \cite{perez1}.

Actually the model in this form suffers from two problems. First a ${\bf 24}$ representation has 24 fermions,
which therefore violates the anomaly\cite{vdbij1, vdbij2} condition. Secondly the fermions are Weyl-fermions. Therefore
the Yukawa  term in eq.(2) is necessarily of the D-type, being symmetric in the interchange
of two types of fermions. As a consequence the $\rho_{32}$ fields stay light. This is not conducive 
to unification, as is easily seen from eq.(4). 

The solution to both problems is simplicity itself. One takes the fermions in the ${\bf 24}$ representation
to be Dirac fermions, instead of Weyl fermions, thereby doubling the number of fermions. This leads to
48 fields, which is divisible by 16. Morever for Dirac fermions there is a second independent
Yukawa interaction of F-type, being antisymmetric in the group indices and proportional to the structure
constants of the group. This leads to an extra contribution to the  mass of the $\rho_{32}$ fields,
which one can move to the unification scale. Furthermore using Dirac fermions one has particle number conservation,
which makes the lightest particle within this multiplet  a candidate for dark matter.

Going from Weyl-fermions to Dirac-fermions the contribution  from the fermions
are doubled.  The one-loop renormalization equations
now read:

\begin{eqnarray}
2\pi(\alpha_1^{-1}-\alpha_U^{-1}) &=& \frac{41}{10} \ln(m_U/m_Z) + \frac{20}{3} \ln(m_U/m_{32})\\
 \nonumber
2\pi(\alpha_2^{-1}-\alpha_U^{-1}) &=& -\frac{1}{2} \ln(m_U/m_Z) - \frac{8}{3} \ln(m_3/m_Z) + 
4 \ln(m_U/m_{32})\\ \nonumber
2\pi(\alpha_3^{-1}-\alpha_U^{-1}) &=& -3 \ln(m_U/m_Z) 
- 4 \ln(m_8/m_Z) + \frac{8}{3} \ln(m_U/m_{32})
\end{eqnarray}

In this equation we take the $\rho_{32}$ field to have a mass at the unification scale.
We therefore leave out the terms $ \ln(m_U/m_{32})$. If we furthermore assume that the triplet
is the only dark matter component of the universe, its mass can be calculated from the 
dark matter abundance. 
Following \cite{cirelli1,cirelli2} we find $m_3 = 1.9\, TeV$. The difference with
\cite{cirelli1,cirelli2} is a factor $1/\sqrt{2}$ in the mass because of the Dirac nature of the fields.

Using as input the central values\cite{pdg} :
\begin{eqnarray}
\alpha_1^{-1} &=& 59.000, \\ \nonumber
\alpha_2^{-1} &=& 29.572,  \\ \nonumber
\alpha_3^{-1} &=& 8.425, 
\end{eqnarray}

one can solve solve the equations and finds:

\begin{eqnarray}
m_8 &=& 6.80 \,\,10^6\, GeV, \\ \nonumber
M_U &=& 4.55\,\, 10^{18}\, GeV, \\ \nonumber
a_U &=& 0.029, 
\end{eqnarray}

\section{Discussion}
The results of the model are quite satisfactory and appear to fulfil the conditions
set out in the introduction.
The spectrum is simple.
The unification scale is quite close to the Planck scale and comes out naturally.

The triplet fields are light. 
The neutral component is $166\, MeV$ lighter than the charged ones, due to
radiative corrections.
This makes the neutral component an excellent candidate
for the dark matter of the universe. If the dark matter would consist of more components, for instance an admixture of scalar singlets, the mass of the triplet fermions would have to be smaller. This would lead to  a further increase in the unification scale towards the Planck mass.
On the other hand lowering $m_{32}$ leads to a lower unification scale; one can even go down to the limits
from proton decay.
The phenomenology of the triplets at the LHC is similar to the phenomenology of light triplet scalars.
 The typical signature of such particles has been studied in \cite{cirelli1,cirelli2,lopez, perez2}. 

The values in eq.(7) are somewhat qualitative and cannot be taken as precision predictions.
 They depend of course on the uncertainties in the determination of
the coupling constants, higher order corrections in the renormalization group, contributions
of scalars, non-perturbative effects from gravity etc. The main conclusion however, that the triplet
can be light, and the unification can take place very close to the Planck scale, will stay true.
 Therefore this model makes two testable predictions.
First one expects a triplet of light Dirac fermions at the LHC. Secondly one expects proton decay
to be absent in experiments in the near future, as the unification scale is near the Planck mass,
which makes the proton lifetime too large. However the last prediction can be avoided when $m_{32}$
is smaller than the unification scale.

Non-supersymmetric unification with ${\bf 24}$'s has been sparsely considered in the 
literature. See for instance \cite{krasnikov:1993}, where however more emphasis was put on the  Higgs field representations.
Some recent models are given in \cite{frigerio, kannike}.
The precise content with a Dirac ${\bf 24}$, which is needed to fulfil the condition
in \cite{vdbij1,vdbij2}  appears not to have been discussed. The presence of a Dirac ${\bf 24}$ would seem to point to an underlying supersymmetry, possibly even N=2 supersymmetry. Indeed a similar spectrum of fermions was found in the  model  in \cite{jones1,jones2},
 which was introduced as an example of a finite unified theory. For a review of finite unified theories
see \cite{zoupanos}. This model could satisfy the
conditions of\cite{vdbij1,vdbij2}, since further Higgs representations come in 
pairs ${\bf 5}$ and ${\bf \bar 5}$. However given the above we would expect supersymmetry
to become relevant only near the Planck scale, which is not unreasonable as
local supersymmetry implies gravity.\\

\noindent {\bf Acknowledgement} I thank Prof. G. Zoupanos for discussions.

\end{document}